# Assigning Satisfaction Values to Constraints: An Algorithm to Solve Dynamic Meta-Constraints

Janet van der Linden

Faculty of Mathematics and Computing, The Open University,
Walton Hall, Milton Keynes MK7 6AA, England
`j.vanderlinden@open.ac.uk`

**Abstract.** The model of Dynamic Meta-Constraints has special activity constraints which can activate other constraints. It also has meta-constraints which range over other constraints. An algorithm is presented in which constraints can be assigned one of five different satisfaction values, which leads to the assignment of domain values to the variables in the CSP. An outline of the model and the algorithm is presented, followed by some initial results for two problems: a simple classic CSP and the Car Configuration Problem. The algorithm is shown to perform few backtracks per solution, but to have overheads in the form of historical records required for the implementation of state**.**

## 1. Introduction

The motivation for the research is the problem of students selecting modules for their studies at a university, which has quite complicated rules as to which combinations are allowed, and in which order modules are to be taken. In particular there are numerous rules like '*if you want to take this module, then this module should be taken first*', or '*if studying this field then at least 7 of the following modules should be taken, but not more than 10*'. A further characteristic is that only parts of the variables and constraints are involved in any particular instance of the problem (i.e. if studying French, then the constraints and variables related to Computing and Business Studies are not applicable).

The above characteristics lead to the area of research where configuration and constraint satisfaction meet, in particular that of Dynamic CSP (DCSP) as defined by Mittal and Falkenhainer[2], and recently re-defined as that of Conditional CSP (CCSP) by Sabin and Freuder[3]. In DCSP or CCSP special constraints are introduced which can activate *variables* to become part of the problem. However, in the current problem domain it is more appropriate to have *constraints* that can be activated if they need to be part of the problem. Secondly, in this problem domain it is important to be able to reason about the *numbers of constraints* that are satisfied or not. This is much in line with recent additions to Constraint Logic Programming, in particular that of the cardinality operator by van Hentenryck[5],[6]. This operator works as a meta-structure, ranging over other constraints. Finally, the user should be able to select a section of the problem, resolve that section, and request for the rest of the problem to be solved in line with the section resolved already. This requires that the problem solving process is modelled as an interactive one, allowing a step-by-step approach. These three ingredients, the *notion of activating constraints*, the *meta-constraints ranging over other constraints* and the *step-by-step* approach to problem solving form the basis of the model of Dynamic Meta-Constraints.

The paper is organised as follows. Firstly, the main features of the model of Dynamic Meta-Constraints are presented. This is followed by a discussion of the algorithm, and the strategy of additional satisfaction values to improve the algorithm's efficiency. It is pointed out that the algorithm may have space requirements problems. Some test results are then presented for a number of problems. The last section concludes with a brief summary and outlines some directions for future work.

## 2. Model of Dynamic Meta-Constraints: Network Representation

For the implementation of the algorithm presented here, the structure as shown in Fig. 1 is used.

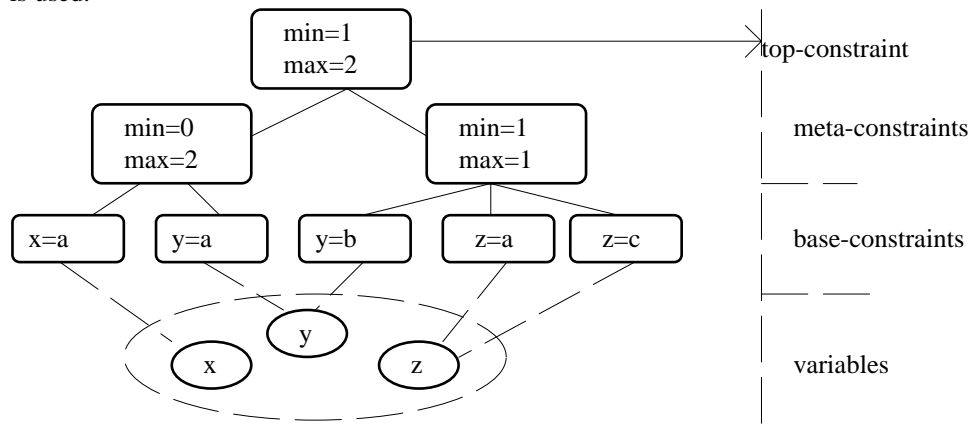

**Fig. 1** Dynamic Meta-Constraints' Network Representation

The variables are stored separate from the constraints and they form an unordered set with no relationships directly between them. The constraints in the problem are linked to each other as in a tree structure, which expresses the fact that they are ordered, and relate to each other as in a parent-child relationship. The constraints and the variables are connected through the base-constraints, which are situated at the bottom of the tree, 'the leaf-nodes'. Because the constraints in these leaf-nodes are connected to variables, the problem as a whole stops being tree-structured, and turns into graph. Therefore, unfortunately, the problem is in essence still like the classic CSP, and belongs to the class of NP-complete problems.

The satisfaction of a base-constraint depends directly on the value assigned to the variable it relates to. Furthermore, a base-constraint cannot itself be a parent to another constraint, and each base-constraint points to one variable only. A meta-constraint has a minimum and a maximum value, and a set of constraints that it ranges over. We say that if at least *min* and at most *max* of the dependent constraints are satisfied, then the meta-constraint is satisfied. The top-constraint represents the problem as a whole.

## 3. Constraints are Active or Inactive and Variables can Remain Unassigned.

Constraints are either active or inactive and at any point during constraint processing, only the currently active constraints participate. This supports the notion that sections of the problem that are not required, are cut off from the problem space through inactivity.

Some variables remain unassigned at the end of constraint processing. The interpretation given to this, is that such variables are not part of the problem currently being solved. This interpretation is analogue to that of DCSP[2]. The two approaches are in fact mirror-images. On the one side is Mittal, with an algorithm that keeps track specifically of the set of *active variables*, in which constraints related to currently inactive variables are simply not considered. While on the other side, we have a careful system of monitoring *which constraints* are active, which forms the basis of a correct implementation of which variables are part of the problem. In the current approach, variables associated with active constraints are said to be active, while other variables are considered inactive.

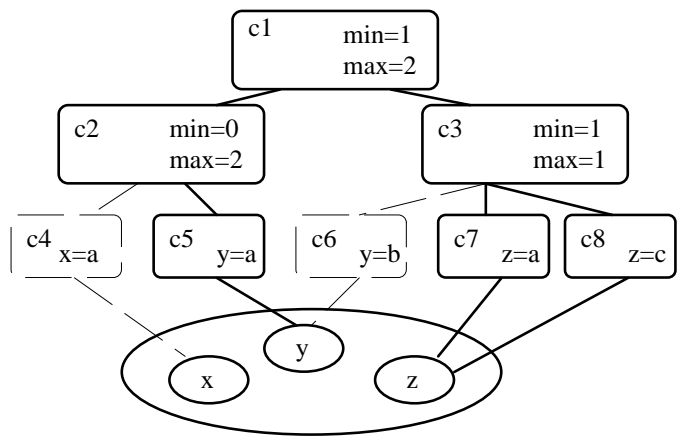

**Fig. 2** Example of a Problem with Active and Inactive Constraints. Activator constraints are omitted from the figure

In Fig. 2 is an example of a constraint network with active and inactive constraints. Inactive constraints are depicted with dotted lines. If the task to satisfy *c3* is executed, this will result in variable Z being assigned, since this is the only active variable associated with constraint *c3* through the constraints *c7* and *c8*. There are two solutions to this request, Z='a' and Z='c'. On the other hand, when executing the task satisfy *c1*, the variables {Y, Z} are both assigned values, while variable X remains unassigned.

## 4  Special Constraints : Activators and Receivers.

Special activator constraints are capable of activating one or several other constraints, upon a condition becoming satisfied. In Fig. 3, at the far right is an example of an activator *c31*, which when the condition Y=a is true, will activate the constraint *c21*, while constraint *c32* is capable of activating several constraints: *c22*, and *c23*. Although no examples are given in this figure, meta-constraints can also be activators.

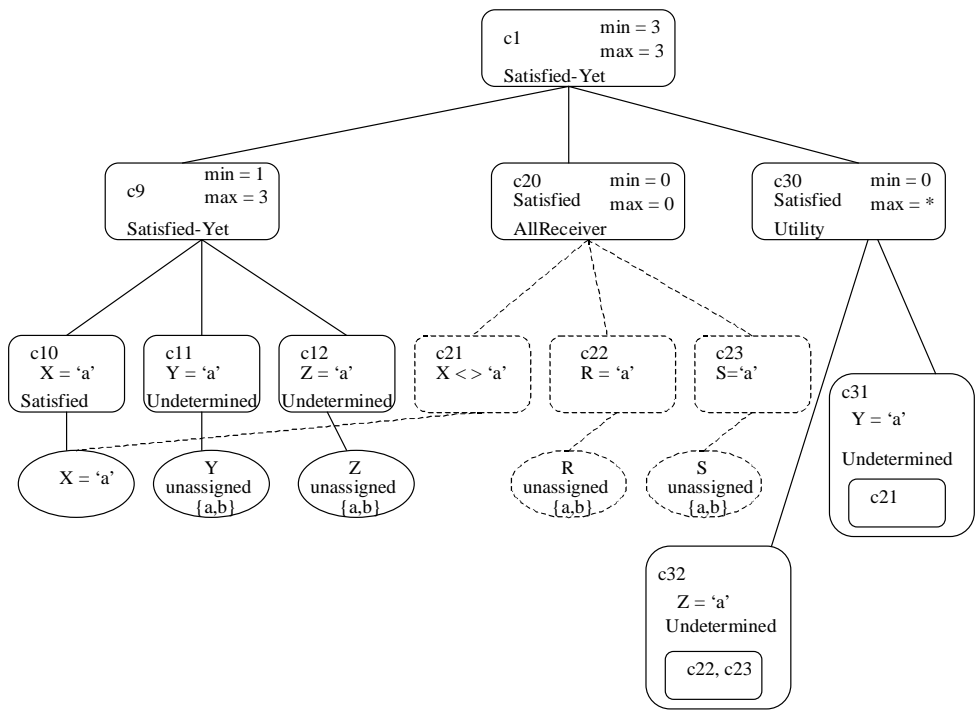

**Fig. 3** An Example of a Constraint Network with Activator Constraints and All-Receiver.

A further type of constraint in this model are Receivers. A Receiver is a meta-constraint, which ranges over other constraints which may or may not become active. In that sense, the meta-constraint is capable of 'receiving' other constraints. A Receiver constraint is satisfied if at least *min* and at most *max* of its *active* dependent constraints are satisfied. Fig. 3 shows *c20* as an example of an AllReceiver. An AllReceiver is a subtype of the Receiver class, and we say that it is satisfied if *all its active children are satisfied*.

Meta-constraints are modelled on the cardinality combinator by Van Hentenryck[5],[6], and to an extent, one could represent problems as demonstrated in this paper using the cardinality combinator, in the CLP scheme. However, the cardinality combinator has not previously been demonstrated as a suitable tool for modelling design problems, such that it models parts and sub parts of the problem, or even the problem as a whole. Furthermore, the cardinality combinator has only been used in the context of regular CSPs, where normal conditions of all variables having to be assigned values, and all constraints having to be satisfied, apply. Regular CSPs do not have activation mechanisms, and don't have the problem that we have here, of not knowing which constraints are the exact dependents of the meta-constraint, nor how many dependants there will be in all. This is why our model has Receiver constraints, which are a more flexible form than the meta-constraint directly modelled on the cardinality combinator. A further difference lies in the implementation of the cardinality combinator compared to the meta-constraints. Van Hentenryck checks whether the constraint store holds information to see if the combinator is satisfied, whereas the current algorithm has a more active approach, and the meta-constraint is used to generate choices, and produce sets of solutions.

## 5. The Algorithm: Propagation in Both Directions.

The algorithm presented here uses propagation through the network going in both directions. The first propagation goes from the variables to the constraints, and is in itself fairly straightforward. When a variable is assigned a value, it will propagate to its associated base-constraints the fact that it has become assigned, resulting in those constraints becoming satisfied or not. The base-constraints then communicate any change in their satisfaction value to their parent meta-constraint, who in turn will notify their parent, etc., until the effects have 'bubbled' up as far as necessary.

The propagation going from the constraints to the variables, can require backtracking to explore all possibilities. A user can request for a constraint to be made satisfied (or unsatisfied). If the constraint is a base-constraint and its associated variable is still unassigned, then the correct value will be assigned to the variable, making the base-constraint satisfied. The task can also fail, in the case where the variable is already instantiated, but with the wrong value.

If the task is to satisfy a meta-constraint then there may be a number of paths to be generated for testing. How many paths depends on the values of *min* and *max*, and the number of dependent constraints of this particular meta-constraint. For example, if the task is to satisfy *c1* in Fig. 2, then the following paths could be generated:

- satisfy *c2* & satisfy *c3*
- satisfy *c2* & unsatisfy *c3*
- unsatisfy *c2* and satisfy *c3*

Each path may lead to numerous other paths. The backtrack algorithm sets out to find all possible paths, or report failure if none exist.

## 6 Five Different Satisfaction Values.

Given that the generation of tasks can grow exponentially, it is important to bring paths to a halt as soon as possible, and to avoid generating paths that are doomed to fail altogether. With just two satisfaction values for constraints, i.e. satisfied and unsatisfied, intelligence that could have guided the search process is lost. Therefore, three additional values are introduced: *Undetermined, Satisfied-Yet,* and *Unsatisfied-Yet*, where the latter two are strictly reserved for meta-constraints. A number of experiments showed that with five satisfaction values, the algorithm's performance improves dramatically [4].

Base-constraints are assigned the value 'Undetermined' when initialised. Later on, they can become Satisfied or Unsatisfied, depending on the values assigned to the variables they are associated with. The meaning of Satisfied and Unsatisfied is that these are fixed values. If a meta-constraint has such a fixed value, it means that in their tree of dependants there are no Undetermined constraints. The fixed-ness also means that, if-constraint *c4* holds value Satisfied, and a request comes to satisfy *c4*, then this will succeed without processing, because this value can be relied on.

The value Satisfied-Yet will be given under the following circumstances:

    **if**    |SAT| + |SAT-YET| $\geq$ C.Min **and**
            |SAT| + |SAT-YET| + |UNDETER| $\leq$ C.Max
    **then return** Satisfied-Yet

where SAT, SAT-YET and UNDETER are sets of dependent constraints which are Satisfied, Satisfied-Yet or Undetermined respectively; where C.Min and C.Max are the constraint's *min* and *max* values; and the condition holds that at least one of the sets SAT-YET, UNSAT-YET or UNDETER is non-empty.

The meaning of Satisfied-Yet is subtle. It means that attempts to make this meta-constraint unsatisfied will fail, but that it is not certain whether the constraint will end up as satisfied. The uncertainty stems from the fact that somewhere in the tree of dependants, there are constraints which are Undetermined. The value Satisfied-Yet allows the insights derived from the *min* and *max* values, and the current state of the dependent constraints, to be used to guide the search process. That is, if a meta-constraint is Satisfied-Yet, it need not be given the task 'Unsatisfy', since this will fail definitely. In a sense this is adding a 'Node-Consistency' flavour to the algorithm, because tasks that are known to fail are as if 'deleted from the constraint's domain'.

The main point is, that as long as there are Undetermined active constraints in a problem, we cannot say with certainty that a solution is found. It may be, that some constraints are making contradictory demands on a variable. Or it may be that an activator constraint jumps into action, adding an entirely new section to the problem to be solved.

## 7    Implementation of 'State'.

A backtrack algorithm needs to ensure it has a sound notion of 'state' - i.e., that when it returns to a previous point in the search tree that the problem behaves exactly the same as it did when it was in that search point before. For this algorithm, the implementation of state is dispersed throughout the problem. That is, the constraints and variables keep historical records of their 'personal' changes throughout the process of solving. This approach requires the notion of a 'global clock', which has a 'tick' incrementation whenever a significant change occurs. Any object that is affected by this event, makes a record of the change. On backtracking, a sweep through the constraints and variables is made, and each object sets itself back in time by looking at its own historical record.

Due to this implementation of state, the algorithm has overheads in the form of the space requirements for the historical records. The space requirements could be exponential, and may be influenced by the type of constraints used, the size of the domains of the variables, the number of constraints and the *min* and *max* values of the meta-constraints.

## 8.    Results: Mackworth's classic CSP

The first problem to report results for, is as introduced by Mackworth[1], and shown in Fig 4. It is an inconsistent CSP, and thus it is not possible to find a combination of values to be assigned to the variables such that all constraints are satisfied.

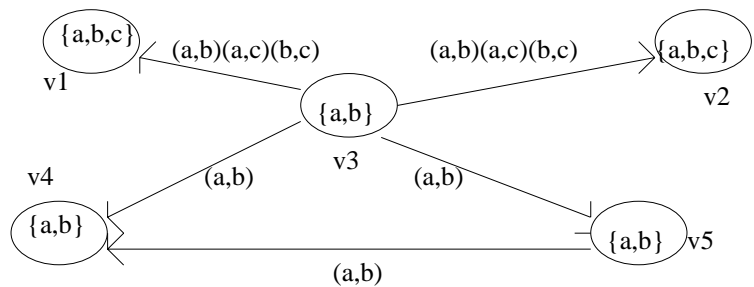

**Fig. 4** Mackworth Classic CSP

The search trees in Table 1 reveal that ordinary backtracking requires 19 assignments, and 12 backtracks to ascertain that no solution is possible for this problem. The algorithm developed for the Dynamic Meta-Constraints performs significantly fewer backtracks and variable assignments. It does perform more constraint checks. More generally, the search trees illustrate the different approaches between the algorithms: ordinary backtracking assigns values to variables and then checks whether the constraints are satisfied, whereas in the new algorithm constraints are made satisfied *by assigning values to variables which are known to be successful*. For example, *v2* is never assigned value 'a', since no constraint indicates that this would be a good choice.

**Table 1.** Search Trees and Results for Classic CSP

| Ordinary Backtrack: | Dynamic Meta-Constraints: |
|---|---|
| v2 = a,  v3 = a,<br>          v3 = b,<br>v2 = b,  v3 = a,  v5 = a<br>                  v5 = b, v4 = a<br>                          v4 = b<br>          v3 = b<br>v2 = c,  v3 = a,  v5 = a,<br>                  v5 = b, v4 = a<br>                          v4 = b<br>          v3 = b,  v5 = a<br>                   v5 = b | v2=c, v3=b<br>v2=b, v3=a, v5=b<br>v2=c, v3=a, v5=b |
| 19 assignments,<br>12 backtracks,<br>18 constraint checks | 8 assignments<br>3 backtracks<br>22 constraint checks |

Arc Consistency algorithms would not make any backtracks or variable assignments when solving the current problem. The procedure REVISE would deplete the domains of some variables of *all* their values, thus indicating that no solution is possible [1].

The results for this very small problem indicate that the efficiency of the Dynamic Meta Constraints approach lies somewhere in between backtracking and Arc Consistency. Obviously, the problem is too small to make any conclusive statements, nor would such statements take into account the issue of the space requirements. Furthermore, the problem is too small and too simple to give proper credit to the new algorithm since it offers no opportunity to demonstrate its more advanced features.

## 9. Results: Mittal's Car Configuration Problem

In the Car Configuration Problem described by Mittal et al[2], and again by Sabin and Freuder[3] special constraints introduce variables which are made part of the problem to be solved. It is therefore closer to the approach of Dynamic Meta-Constraints than the previous problem. This particular problem consists of configuring cars with accessories according to some level of luxury: *'standard, deluxe and luxury'*. Depending on these the car will or will not be fitted with air-conditioner, sun-roofs etc..

| Variable | Domain | |
|---|---|---|
| package | {luxury, deluxe, standard} | *initial var*[1] |
| frame | {convertible, sedan, hatchback} | *initial var* |
| engine | {small, med, large} | *initial var* |
| battery | {small, med, large} | |
| sunroof | {sr1, sr2} | |
| airconditioner | {ac1, ac2} | |
| glass | {tinted, non-tinted} | |
| opener | {auto, manual} | |

**Activity Constraints**[2]
1. Package = luxury  RV⇒ Sunroof
2. Package = luxury RV⇒ Airconditioner
3. Package = deluxe RV⇒  Sunroof
4. Sunroof = sr2 RV⇒ Opener
5. Sunroof = sr1 RV⇒ Airconditioner
6. Sunroof  ARV⇒  Glass
7. Engine ARV⇒ Battery
8. Opener ARV⇒ Sunroof
9. Glass ARV⇒ Sunroof
10. Sunroof = sr1 RN⇒ Opener
11. Frame = convertible RN⇒ Sunroof
12. Battery = small & Engine = small RN⇒ Airconditioner

**Compatibility Constraints**
13. Package = standard → AirConditioner ≠ ac2
14. Package = luxury  → AirConditioner ≠ ac1
15. Package = standard → Frame ≠ convertible
16. Opener = auto & AirConditioner = ac1 → Battery = med
17. Opener = auto & AirConditioner = ac2 → Battery = large
18. Sunroof = sr1 & AirConditioner = ac2 → Glass ≠ tinted

---

[1] Initial variables are those variables that definitively need to be assigned a value. The remainder of the variables need to be activated through activity constraints before values are assigned to them.

[2] In Mittal's notation RV stands for 'Requires Variable', ARV for 'Always Requires Variable' and RN for 'Requires Not'. The difference between ARV and RV is that with ARV a variable is required independent of the value assigned to the current variable.

Table 2. Results for Mittal's Car Configuration problem, with 64 constraints and 8 variables. Column 3 presents figures for requests to make top-constraint satisfied, finding all 288 solutions, and column 4 for finding the first solution. Column 5 and 6 show figures after constraint stating that *P=deluxe* is made satisfied followed by a request to make the top-constraint satisfied, with now 153 possible solutions found in column 5, and the first solution in column 6

|  |  |  |  | *P=deluxe* | *P=deluxe* |
|---|---|---|---|---|---|
|  |  | AllSols | FirstSol | AllSols | FirstSol |
| Backtracks |  | 290 | 1 | 153 | 1 |
| Assignments |  | 660 | 4 | 291 | 6 |
| constraint checks |  | 3437 | 26 | 1281 | 37 |
|  |  |  |  |  |  |
| constraint history | min | 1 | 1 | 2 | 1 |
|  | max | 576 | 6 | 308 | 6 |
|  | average | 138.76 | 2.92 | 61.35 | 2.75 |

Table 2 shows that just 290 backtracks need to be made to find all 288 solutions (column 3), or 153 backtracks to find all 153 solutions. This gives an average of 1.0 backtrack per solution. It appears that the only reason the algorithm has to backtrack, is when it has found a solution, and thus avoids going down any unnecessary paths.

Unfortunately there are no figures available from Mittal's implementation, so a comparison is not possible. Neither is it possible to compare the figures from Table 2 with an ordinary backtrack or Arc Consistency algorithm, since these would not be able to solve a problem modelled in this fashion

A drawback of the algorithm is the average number of historical records accumulated during constraint processing which, particularly given the size of this problem, is very high. However, results reported in [4] suggest that compared to other problems, Mittal's Car Configuration Problem may be unusual in having such a high average. It would appear that the 'problem texture' is of great importance and that Mittal's problem is unusually dense, with tight, close interaction between the variables and constraints.

## 8  Conclusion

By propagating values from the constraints to the variables, variables are being assigned 'correct' values, rather than values which then need to be tested. Even though the meta-constraints can generate large numbers of paths, it was found that the algorithm had a remarkable near one-to-one ratio for the number of backtracks required per solution. The additional satisfaction values that were introduced, provide a form of consistency checking, which reduces the search space considerably. Future work will aim to reduce the accumulation of historical records, and focus on the development of heuristics to guide the selection of constraints during the process of constraint solving.

# References

1. Mackworth, A. K.: Consistency in Networks of Relations. In: Artificial Intelligence, 8 (1977) pp 99-118
2. Mittal, S., Falkenhainer, B.: Dynamic Constraint Satisfaction Problems. In: Proceedings AAAI '90, Boston, MA (1990)
3. Sabin, M., Freuder, E. C.: Detecting and Resolving Inconsistency and Redundancy in Conditional Constraint Satisfaction Problems. In: Configuration, Papers from the AAAI Workshop, WS-99-05, AAAI Press, Menlo Park, California (1999) pp 90-94
4. Van Der Linden, A. S. J.: Dynamic Meta-Constraints: An Approach to Dealing with Non-Standard Constraint Satisfaction Problems. PhD thesis, Oxford Brookes University, Oxford, England (2000)
5. Van Hentenryck, P., Deville Y.: The Cardinality Operator: A New Logical Connective for Constraint Logic Programming. In: Furakawa, K.(ed.): Proc. of the 8th International Conference, France, (1991) pp 745-759
6. Van Hentenryck, P., Simonis, H., Dincbas, M.: Constraint Satisfaction using Constraint Logic Programming. In: Artificial Intelligence 58 (1992), pp 113-159